\documentclass[11pt]{article}

\usepackage[final]{acl}
\usepackage{amsmath,amssymb}
\usepackage{pgfplots}
\usepackage{multirow}
\usepackage{graphicx}
\usepackage{placeins}
\usetikzlibrary{patterns} 
\usepackage{comment}
\usepackage{times}
\usepackage{inconsolata}
\pgfplotsset{compat=1.17} 
\AtBeginDocument{%
  }

\usepackage{xcolor}
\definecolor{myblue}{RGB}{31,119,180}   
\definecolor{myred}{RGB}{214,39,40}       
\definecolor{mygreen}{RGB}{44,160,44}      
\definecolor{myorange}{RGB}{255,140,0} 

\pgfdeclarepatternformonly[pattern color=myblue]{sparse north east lines}
  {\pgfpoint{0pt}{0pt}}{\pgfpoint{6pt}{6pt}}{\pgfpoint{6pt}{6pt}}%
  {
    \pgfsetlinewidth{0.6pt}  
    \pgfpathmoveto{\pgfpoint{0pt}{0pt}}
    \pgfpathlineto{\pgfpoint{6pt}{6pt}}
    \pgfusepath{stroke}
  }

\pgfdeclarepatternformonly[pattern color=mygreen]{sparse horizontal lines}
  {\pgfpoint{0pt}{0pt}}{\pgfpoint{6pt}{6pt}}{\pgfpoint{6pt}{6pt}}%
  {
    \pgfsetlinewidth{0.4pt}
    \pgfpathmoveto{\pgfpoint{0pt}{3pt}}
    \pgfpathlineto{\pgfpoint{6pt}{3pt}}
    \pgfusepath{stroke}
  }

\pgfdeclarepatternformonly[pattern color=myred]{sparse south west lines}
  {\pgfpoint{0pt}{0pt}}
  {\pgfpoint{6pt}{6pt}}
  {\pgfpoint{6pt}{6pt}}
  {%
    \pgfsetlinewidth{0.6pt}
    \pgfpathmoveto{\pgfpoint{0pt}{6pt}}
    \pgfpathlineto{\pgfpoint{6pt}{0pt}}
    \pgfusepath{stroke}
  }

\pgfdeclarepatternformonly[pattern color=myred]{sparse vertical lines}
  {\pgfpoint{0pt}{0pt}}{\pgfpoint{6pt}{6pt}}{\pgfpoint{6pt}{6pt}}%
  {
    \pgfsetlinewidth{0.4pt}
    \pgfpathmoveto{\pgfpoint{3pt}{0pt}}
    \pgfpathlineto{\pgfpoint{3pt}{6pt}}
    \pgfusepath{stroke}
  }

\author{
 \textbf{Kayhan Behdin\textsuperscript{1,3,5}},
 \textbf{Ata Fatahibaarzi \textsuperscript{1,3,5}},
 \textbf{Qingquan Song\textsuperscript{2,3,5}},
 \textbf{Yun Dai \textsuperscript{2,3,5}},
\\
 \textbf{Aman Gupta \textsuperscript{2,3,5}},
 \textbf{Zhipeng Wang \textsuperscript{$^\ast$,1,3}},
 \textbf{Shao Tang \textsuperscript{1,3}},
 \textbf{Hejian Sang \textsuperscript{1,3}},
\\
 \textbf{Gregory Dexter \textsuperscript{1}},
 \textbf{Sirou Zhu \textsuperscript{1}},
 \textbf{Siyu Zhu\textsuperscript{1}},
 \textbf{Tejas Dharamsi\textsuperscript{1}},
 \\
 \textbf{Vignesh Kothapalli\textsuperscript{2}},
 \textbf{Zhoutong Fu\textsuperscript{1}},
 \textbf{Yihan Cao\textsuperscript{1}},
 \textbf{Pin-Lun Hsu\textsuperscript{2}},
 \\
 \textbf{Fedor Borisyuk\textsuperscript{1}},
 \textbf{Natesh Pillai\textsuperscript{1}},
 \textbf{Luke Simon\textsuperscript{2}},
  \textbf{Rahul Mazumder \textsuperscript{1,3,4}}
\\
\\
 \textsuperscript{1} LinkedIn,
 \textsuperscript{2} Work done at LinkedIn,  
 \textsuperscript{3} LinkedIn LLM Efficiency Core Team, 
 \textsuperscript{4} MIT \\
 \small {\textsuperscript{5} Authors contributed equally to this research}, 
 \small{
   \textsuperscript{$\ast$}
   \textbf{Correspondence:} \href{mailto:zhipwang@linkedin.com}{zhipwang@linkedin.com}
 }}

\title{Scaling Down, Serving Fast: Compressing and Deploying Efficient LLMs for Recommendation Systems}

\begin{document}

\maketitle
\begin{abstract}
Large language models (LLMs) have demonstrated remarkable performance across a wide range of industrial applications, from search and recommendation systems to generative tasks. Although scaling laws indicate that larger models generally yield better generalization and performance, their substantial computational requirements often render them impractical for many real-world scenarios at scale. In this paper, we present a comprehensive set of insights for training and deploying small language models (SLMs) that deliver high performance for a variety of industry use cases. We focus on two key techniques: (1) knowledge distillation and (2) model compression via structured pruning and quantization. These approaches enable SLMs to retain much of the quality of their larger counterparts while significantly reducing training/serving costs and latency. We detail the impact of these techniques on a variety of use cases in a large professional social network platform and share deployment lessons, including hardware optimization strategies that improve speed and throughput for both predictive and reasoning-based applications in Recommendation Systems.
\end{abstract}

\section{Introduction}

Large language models (LLMs)~\cite{dubey2024llama, jiang2023mistral, team2023gemini, liu2024deepseek} have ushered in a new era in artificial intelligence and machine learning, driving significant improvements in deployed systems across various industries.

LLMs suitable for real-world applications come in diverse forms, differing in size (ranging from hundreds of millions to hundreds of billions of parameters), architectural design (e.g., encoder-based models like BERT~\cite{devlin2018bert} versus decoder-based models like GPT-3~\cite{brown2020language}), and training paradigms (such as pre-training, instruction tuning, or test-time computation~\cite{dubey2024llama,team2023gemini,jiang2023mistral,mueller2023meta,liu2024deepseek,guo2025deepseek}).

Specifically for the Social Network Platforms, LLMs are heavily leveraged and deployed for a multitude of applications: 1.~Semantic Search (e.g., embedding generation \cite{wang2022text, wang2023improving} and semantic ranking/matching~\cite{qin2023large}); 2.~Recommendation Systems (RecSys), specifically Retrieval and Ranking~\cite{zhao2024dense}~\cite{li2023agent4ranking,firooz2025360brew,li2023text}; 3.~Generative Use cases, such as chatbots, assistants, image generators, etc.~\cite{achiam2023gpt, dam2024complete, ramesh2022hierarchical}. 

Furthermore, scaling laws for LLMs have established a strong correlation between model size, validation loss, and downstream task performance~\cite{kaplan2020scaling, hoffmann2022training, raffel2020exploring, wei2022chain}. As a result, increasing the size of the model is often one of the most effective strategies to enhance performance. Modern LLMs, particularly autoregressive decoder-only models, have expanded to hundreds of billions of parameters.

Although large LLMs exhibit extraordinary performance, the deployment of such large models incurs substantial infrastructure costs, especially for latency- or throughput-sensitive tasks in Recommendation Systems (RecSys). However, both academia and industry have developed strategies for creating and deploying efficient small language models (SLMs). Here, we primarily focus on methods that leverage an existing internally trained large LLM to create an efficient SLM that largely maintains the original model's accuracy. Approaches to achieve this include white-box or black-box distillation~\cite{hinton2015distilling, gu2024minillm, jin2021modality, agarwal2024policy, tunstall2023zephyr}, compression techniques such as quantization~\cite{frantar2022gptq, behdin2023quantease} and sparsification~\cite{frantar2023sparsegpt, meng2024osscar, sun2023simple, meng2024alps,oberta}.

In this work, we present a suite of insights from the training and deployment of various efficient SLMs in production at a large-scale professional social networking company. We address a wide array of predictive and generative use cases in RecSys (including ranking, recommendations, and reasoning), with inference performance and latency constraints in serving as key considerations. Our contributions are as follows.
\begin{itemize}
    \item We discuss several large-scale RecSys use cases for which language models are useful.
    \item For these use cases, we explore techniques for developing tailored SLMs, with a focus on knowledge distillation and model compression methods such as quantization and structured pruning.
    \item We discuss inference, latency, and other serving considerations, offering insights into the infrastructure required to reliably deploy SLMs in high-throughput or low-latency production environments, and share practical lessons from our real-world deployments.
\end{itemize}

\section{Preliminaries}
\label{section:prelim}

\textbf{Training details} We consider models ranging in size from a billion to $\sim 100\text{B}$ parameters. For all use cases, we appropriately tune the learning rate, learning rate warmup schedule and decay, as well as weight decay. Context length varies from a few hundred tokens to up to 32k, depending on the use case. We provide use-case specific details at the appropriate place.

\noindent \textbf{Prompt Structures} For predictive tasks, we are mainly interested in ranking use cases. Hence, the use of decoder-based LLMs is prefill-dominant. For generative and reasoning based-tasks, we are also interested in decoding latencies. We relegate the details of the prompt structures to subsections for the specific use cases.

\noindent \textbf{Quality metrics} We use different accuracy measures across tasks. For predictive tasks, we use area under the curve (AUC). For generative tasks, we rely on validation loss and task-specific metrics.

\noindent \textbf{Foundational model for RecSys} We base our experiments on an internal foundation model (FM) trained using text, primarily for the purpose of ranking and recommendations
~\cite{firooz2025360brew}. The FM is a Mixture-of-Experts (MoE) model with an architecture motivated by the Mixtral family~\citep{jiang2024mixtral}. In particular, each expert is initialized based on a Llama 3.1 8B Instruct model~\citep{dubey2024llama}, with 16 experts in total (4 active per token). 
The FM is trained to approximate the following distribution for a large variety of recommendation tasks where users interact with items:  
\begin{equation}
    P (m, (e_1, t_1),...,(e_T, t_T)),
\end{equation}
where $m$ represents a user, and each pair $(e_t, i_t)$ for $t = 1, ..., T$ represents the user's interaction $i_t$ (like or click or equivalent) with an entity $e_t$ (such as post on a social media platform) . As mentioned before, the featurization is performed purely via text, allowing the FM to effectively generalize across heterogeneous tasks. The model is then used to estimate the following probabilities for future interactions with entities:

\begin{multline}
    P(i_t, i_{t+1}, \dots \mid \text{Task instruction}, m, \\
    (e_1, i_1), \dots, (e_{t-1}, i_{t-1}), e_t, e_{t+1}, \dots)
\end{multline}

Text-based featurization makes decoder-only LLMs an attractive option for training this model jointly on a large and varied of recommendation tasks. The FM uses single-token generation for pointwise ranking and probability estimation. The reader is encouraged to peruse the paper~\citet{firooz2025360brew} for more details about the FM.

Due to the large size of the FM---which contains over 100 billion parameters---serving it online for latency-sensitive applications is challenging. In this work, we present experiments demonstrating how we achieved a more than 20$\times$ reduction in model size, enabling the online serving of a compressed version of the FM with only a modest loss in accuracy.

\begin{figure*}[hbpt]
    \centering
    \includegraphics[width=0.9\textwidth]{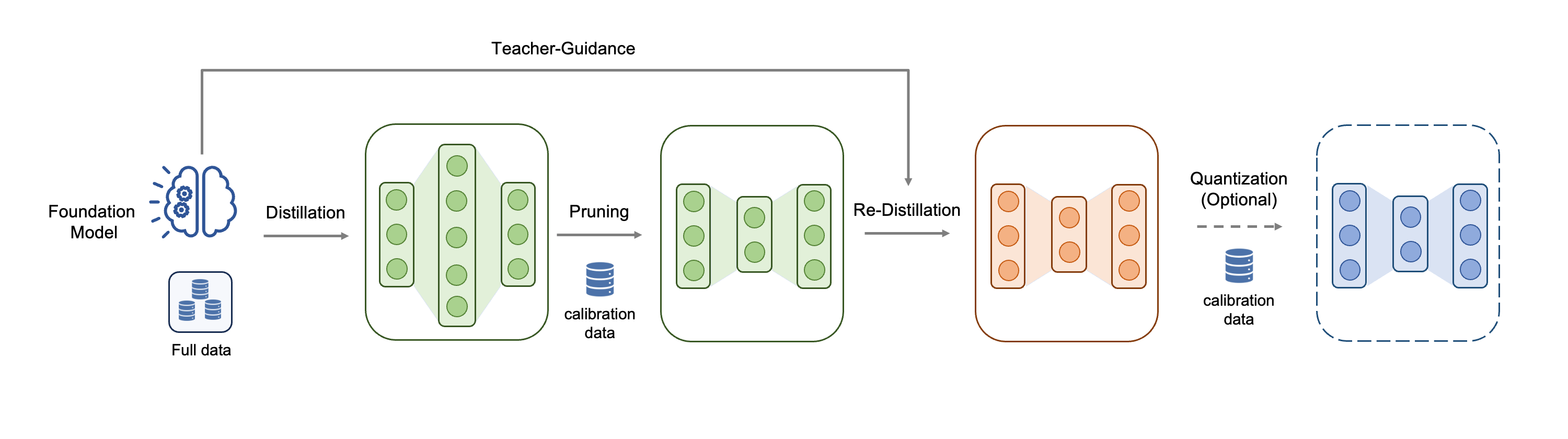} 
    \caption{Overview of the process of creating SLMs via distillation and compression.}
    \label{fig:mental_model}
\end{figure*}
\section{Methodology}
We apply these methods to the following real-world use cases in RecSys, with detailed methodology described in 
Appendix~\ref{section:methods}.

\begin{itemize}
    \item \textbf{SLM for predictive tasks obtained through distillation and  pruning} - We leverage a foundation model (FM) for ranking and recommendations~\cite{firooz2025360brew, li2023text} and leverage distillation and pruning to create an SLM that is efficient for serving latency-sensitive use cases. The final SLM we create is more than $20\times$ smaller without any appreciable loss in quality.
    \item \textbf{SLM for reasoning task obtained through distillation} - We leverage various flavors of KD to compress a latency-sensitive reasoning model by more than $5\times$ with comparable quality. 
\end{itemize}

Our approach, illustrated in Figure~\ref{fig:mental_model}, proceeds in three stages: 1)~\textbf{Distillation} on the full model. 2)~\textbf{One-shot structured pruning}\footnote{``One-shot'' indicates pruning without re-training.} to significantly reduce the model size. 3)~\textbf{Re-distillation} of the pruned model to recover generalization capabilities.
All compression is performed in a post-training setting using data from various recommendation tasks (see Section~\ref{section:methods} for details on methods). We optionally also quantize the model (details in Section~\ref{section:deployment}). We measure quality by reporting the AUC on the test sets of in-domain tasks, computing AUC per task and then averaging across tasks.

\section{Experimental Results for Reasoning Tasks}

We detail experiments for a use case that requires generating reasoning for various input prompts.

We investigate several distillation methods to produce a 1.5B model initialized from Qwen-2.5-1.5B-Instruct \cite{yang2024qwen2}. The teacher models are obtained by training various sizes of Qwen-2.5 Instruct models using SFT with identical hyperparameters. Performance is primarily measured by validation loss, with lower values indicating higher accuracy (see Table~\ref{tab:distillation_results}). 

In a single-stage, word-level training setup, the Forward KL (FKL, $\beta=0$) loss achieves the lowest validation loss, outperforming the Jensen–Shannon Divergence (JSD, $\beta=0.5$), SFT, and Reverse KL (RKL, $\beta=1.0$) approaches. Notably, a larger teacher model does not necessarily lead to superior outcomes; for example, a 3B$\rightarrow$1.5B distillation can outperform a 7B$\rightarrow$1.5B configuration.

In contrast, a two-stage approach---an initial word-level distillation phase followed by on-policy training with the FKL loss (oFKL)---consistently yields better performance than single-stage training alone. Our experiments show that initializing the second stage with the best checkpoint from the first stage (e.g., an FKL-distilled model) is more effective than starting from either an SFT model or the original, non-SFT model. Additionally, an on-policy sampling fraction of 1.0 maximizes accuracy, although a fraction of 0.5 (fr=0.5) may be preferred when faster training is desired. We also observe that generating approximately 300 tokens in this task (denoted as tk in Table~\ref{tab:distillation_results}) during on-policy updates strikes an optimal balance between performance and efficiency, outperforming both 200-token and 400-token alternatives, while a generation temperature in the range of 0.8--0.9 is generally most effective.

Interestingly, the student model can sometimes surpass the teacher’s performance. In the two-stage training paradigm, employing larger teacher models (e.g., 14B rather than 7B or 3B) appears to provide additional benefits for the 1.5B student, although this trend is less pronounced in the single-stage FKL-only training.

Overall, these findings underscore the effectiveness of a two-stage training strategy: an initial supervised fine-tuning phase establishes a strong generative foundation, and subsequent on-policy distillation refines the model’s capabilities, leading to improved generalization and performance.

\begin{table}[htbp]
\scriptsize
\centering
\begin{tabular}{c c c}
\hline\hline
\textbf{Ablation} & \textbf{Training Method} & \textbf{Val Loss} \\
\hline\hline
\multirow{4}{*}{Baseline} & SFT & 0.2236 \\
& 3B-SFT & 0.2081 \\
& 7B-SFT & 0.1941 \\
& 14B-SFT & 0.1771\\
\hline
\multirow{4}{*}{Single Stage}
& FKL & 0.2045 \\
& FKL (3B) & \textbf{0.2015} \\
& JSD ($\beta=0.5$) & 0.2143 \\
& RKL & 0.2333 \\
& oFKL & 0.2107 \\
\hline
\multirow{7}{*}{Two Stages} & SFT-SFT & 0.2295 \\
& SFT-oFKL & 0.1982 \\
& FKL-oFKL & 0.1939 \\
& FKL-oFKL (tk=200) & 0.1910 \\
& FKL-oFKL (tk=300) & \textbf{0.1894} \\
& FKL-oFKL (tk=400) & 0.1910 \\
& FKL-oFKL (tk=300, fr=0.5) & 0.1917 \\
\hline
\multirow{4}{*}{Different Teachers} & FKL (3B)-oFKL (3B) & 0.1954 \\
& FKL (3B)-oFKL (7B) & 0.1918 \\
& FKL (7B)-oFKL (7B) & 0.1894 \\
& FKL (14B)-oFKL (14B) & \textbf{0.1863} \\
\hline\hline
\end{tabular}
\caption{Validation losses for various training methods and ablations. Unless explicitly specified (e.g., FKL (3B) indicates distillation using the 3B-SFT model as the teacher), the default teacher is the 7B-SFT model.}
\label{tab:distillation_results}
\end{table}

To further study the effectiveness of our distillation recipes, we apply them to other open source models. In particular, we use the Qwen3~\citep{yang2025qwen3} model family. On an internal reasoning task, using the same recipes discussed above, we distill a Qwen3 14B model into a Qwen3 4B student model that matches the same accuracy without any performance drops. In addition, using the OpenThoughts reasoning dataset \cite{guha2025openthoughtsdatarecipesreasoning}, we distill the Qwen3 32B reasoning model into Qwen3 8B and 4B student models. We evaluate the student models on AIME 2024 (and 2025) \cite{lighteval}  benchmarks. Using our recopies, we improve the student model's performance by more than $20\%$ for the 8B student model and $15\%$ for the 4B student model compared to their initial base model.

\begin{table}[t]
\scriptsize
\centering
\begin{tabular}{lc}
\hline\hline
\textbf{Model} & \textbf{AUC Delta (\%)} \\
\hline\hline
8B Distilled Model  & -\\ 
6.4B Pruned Model (20\%) + SFT   &-0.47\%  \\
6.4B Pruned Model (20\%) + Distillation   &-0.06\%  \\
\hline\hline
\end{tabular}
\caption{Distillation vs. SFT for post-pruning retraining.}
\label{table:prune_sft_vs_distil}
\label{tab:sft_kd_for_prune}
\end{table}

\begin{table}[t]
\scriptsize
\centering
\begin{tabular}{lcc}
\hline\hline
\textbf{Model}  & \textbf{AUC Delta (\%)} \\
\hline\hline
8B Model    &  - \\
6.8B Pruned Model    & 0.0\% \\
6.4B Pruned Model   &  -1.33\% \\
6.0B Pruned Model   &  -1.72\% \\

\hline\hline
\end{tabular}
\caption{Evaluation of the 8B-parameter model post-SFT and its pruned variants, focusing on MLP pruning performed in a one-shot manner (i.e., no retraining after pruning).}
\label{tab:amount_pruning}
\end{table}

\begin{table}[t]
\scriptsize
\centering
\begin{tabular}{lcc}
\hline\hline
\textbf{Model} & \textbf{\#Params} & \textbf{AUC Delta (\%)} \\
\hline\hline
3B (Distilled from FM)         & 3B &  -\\
MLP Prune + Distill    & 2.4B &  -0.12\%\\
MLP Prune + Distill (Gradual)   & 2.4B & 0.03\% \\

\hline\hline
\end{tabular}
\caption{Comparison of one-step vs gradual pruning for a 3B model distilled from the FM.}
\label{tab:gradual_pruning}
\end{table}

\begin{table}[t]
\scriptsize
\centering
\begin{tabular}{lcc}
\hline\hline
\textbf{Model} & \textbf{\#Params} & \textbf{AUC Delta (\%)} \\
\hline\hline
MLP Pruning       & 2.4B &  -\\
$\wedge$ + Attention Pruning    & 2.1B &  -1.07\%\\
$\wedge$ + Distillation  & 2.1B & 0.02\% \\

\hline\hline
\end{tabular}
\caption{Results for attention pruning. We consider the 2.4B gradual pruning model from Table~\ref{tab:gradual_pruning} as the base. The second row shows the result for one-shot attention pruning, while the last row shows the results after performing distillation.}
\label{tab:mlp_vs_attn_pruning}
\end{table}

\section{Experimental Results for Predictive Tasks}
\subsection{Knowledge Distillation Findings}

To evaluate how knowledge distillation (KD) affects model generalization, we compare it against standard supervised fine-tuning (SFT). We focus on the task performance retention of the distilled (or fine-tuned) student models relative to the original foundation model (FM). Specifically, we use Llama-3.1-8B-Instruct and Llama-3.2-3B-Instruct~\cite{dubey2024llama} as our student models. These models offer strong performance while remaining sufficiently compact for throughput- and latency-sensitive environments. In both KD and SFT, responses are generated from the ground-truth action-prediction labels.
For KD, the per-token loss is a weighted combination of: 90\% forward KL divergence between teacher and student logits, and 10\% cross-entropy loss with the ground-truth labels.

Additionally, we include an extra 5\% loss contribution computed over the entire sequence (including the prompt), normalized by its token count. Thus, 95\% of the loss is computed solely on action prediction tokens (i.e., yes/no token), while the remaining 5\% is computed over the prompt tokens. This allows the model to maintain its foundation model knowledge, helping generalization to unseen tasks.

\begin{figure}[t]
    \centering
\begin{tikzpicture}
    \begin{axis}[
        ybar,
        bar width=0.6cm,
        width=6cm,
        height=6cm,
        ylabel={AUC Delta (\%)},
        ylabel style={font=\small},
        xtick={8B-KD, 8B-SFT, 3B-KD, 3B-SFT},
        symbolic x coords={8B-KD, 8B-SFT, 3B-KD, 3B-SFT},
        scaled y ticks=false,
        enlarge x limits=0.2,
        ymin=-1.4,
        ymax=0.1,
        nodes near coords={\pgfmathprintnumber[fixed, precision=2]{\pgfplotspointmeta}},
        every node near coord/.append style={font=\footnotesize},
        x tick label style={font=\small, rotate=45, anchor=east},
        title style={font=\small},
    ]
    \draw [thin, dashed] (rel axis cs:0,0.934) -- (rel axis cs:1,0.934);
    
    \addplot[fill=myblue!50!white, bar shift=0pt,draw=black!70] coordinates {(8B-KD, -0.06)};
    \addplot[fill=myorange!50!white, bar shift=0pt,draw=black!70] coordinates {(8B-SFT, -0.62)};
    \addplot[fill=myblue!50!white, bar shift=0pt,draw=black!70] coordinates {(3B-KD, -0.15)};
    \addplot[fill=myorange!50!white, bar shift=0pt,draw=black!70] coordinates {(3B-SFT, -1.21)};
    \end{axis}
\end{tikzpicture}
    \caption{Comparison of Distillation and SFT on the Foundation Model. Knowledge distillation consistently outperforms SFT by effectively leveraging teacher supervision to preserve and enhance performance.}
    \label{fig:distillation_sft_compare}
\end{figure}
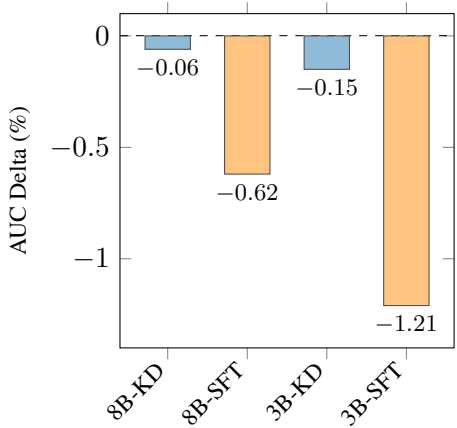

Each KD and SFT configuration undergoes hyperparameter tuning (e.g., peak learning rate, warmup schedule, decay schedule, and weight decay). Figure~\ref{fig:distillation_sft_compare} summarizes the results by reporting the AUC delta relative to the original FM (which achieves the best performance). The main observations are:
\begin{itemize}
    \item \textbf{SFT}: The 3B and 8B student models fine-tuned only with SFT underperform compared to the FM, which is expected given their smaller size and post-training. The performance drop for the 3B model ($-1.21\%$) is larger than that of the 8B model ($-0.62\%$).
    \item \textbf{KD}: Using logit supervision from the FM consistently preserves task performance better than SFT. The 8B-KD model shows a minor AUC drop of $-0.06\%$ (compared to $-0.62\%$ for 8B-SFT), while the 3B-KD model (-0.15\%) substantially mitigates the loss relative to 3B-SFT ($-1.21\%$). These results demonstrate the effectiveness of KD for transferring knowledge.
\end{itemize}

\subsection{Post-training Compression Findings}

After obtaining the distilled student models, we apply structured pruning and on-the-fly FP8 quantization to further compress the model to meet our serving latency requirements (steps outlined in Figure \ref{fig:mental_model}). Since the distilled models (DMs) are decoder-only transformers, our approach focuses on applying structured pruning to remove redundant MLP up/down projection neurons, as well as attention heads in each transformer layer while preserving its capabilities on in-domain ranking tasks. Since an off-the-shelf application of one-shot pruning can result in a loss in model quality (utility), we apply targeted fine-tuning after each pruning step to ensure that the model adapts to its reduced size without significant performance loss. To this end, we again leverage knowledge distillation to bridge the gap between the original and pruned models, transferring key insights and ensuring the pruned version closely aligns with the outputs of the original foundational model. 

For structured pruning, we leverage the OSSCAR algorithm~\cite{meng2024osscar} (see Section~\ref{sec:appendix-compression} for details on methodology). We now discuss the details of various ablations that we conducted towards structured pruning.

\noindent \textbf{Effect of SFT vs. distillation}
As illustrated in Figure~\ref{fig:mental_model}, we employ OSSCAR to prune the distilled model in a layerwise fashion. After pruning, we use either SFT or KD to restore any lost generalization, with the unpruned distilled model serving as the teacher. To demonstrate the effectiveness of each approach, we examine an 8B distilled model and its 6.4B pruned counterpart (i.e., 20\% pruning of the MLP layers). Table~\ref{table:prune_sft_vs_distil} presents the results. Consistent with the findings for pure distillation, the pruned model benefits significantly more from KD than from SFT. Results for the 3B model (pruned to 2.4B) mirror this trend and are omitted for brevity.

While SFT offers a more straightforward optimization path, distillation provides additional flexibility by leveraging teacher--student training to refine model weights more effectively. In practice, the choice of distillation algorithms and associated losses (e.g., forward KL combined with SFT loss) may vary depending on data availability, computational constraints, and the chosen pruning ratio. Nevertheless, in most cases, including a forward KL term proves highly beneficial in counteracting the performance drop associated with pruning.

Pruning the 8B and 3B distilled models down to 6.4B and 2.4B, respectively, can yield further improvements in serving efficiency. Additional details on deployment are provided in Section~\ref{section:deployment}.

\noindent \textbf{Effect of degree and schedule of pruning}
We next investigate how varying the pruning ratio impacts downstream task accuracy. Table~\ref{tab:amount_pruning} reports one-shot pruning results (i.e., no SFT or KD is applied post-pruning) on an 8B model that has undergone an SFT stage (based on Llama-3.1-
8B-Instruct) comparable to the foundation model. As expected, more aggressive pruning reduces model size but also significantly degrades task accuracy. Notably, small pruning ratios (e.g., from 8B to 6.8B) have minimal effect on performance, while heavier pruning leads to larger accuracy drops. However, as shown in Table~\ref{tab:sft_kd_for_prune}, such performance losses can be mitigated by applying distillation.

We also explore \emph{gradual pruning}~\cite{benbaki2023fast,meng2024falcon} in which the model is pruned in multiple steps, with knowledge distillation after each pruning. Table~\ref{tab:gradual_pruning} summarizes results for pruning a distilled 3B model to 2.4B (a 37.5\% MLP sparsity) either in a single step or in two smaller steps (removing 1536 hidden neurons in each step, for a total of 3072). The gradual approach recovers model AUC better than the single-step approach, achieving near-lossless compression from 3B to 2.4B.

Finally, to study attention-head pruning, we take the 2.4B model obtained by gradual MLP pruning and prune half of its attention heads via OSSCAR in a one-shot manner, followed by KD. Table~\ref{tab:mlp_vs_attn_pruning} shows that one-shot attention pruning incurs a modest quality loss, but the subsequent KD phase restores the model to AUC parity with the 2.4B baseline.

We also analyze the effect of calibration data for pruning and results are summarized in Appendix~\ref{appendix:calibration_data_pruning}

\section{Deployment}
\label{section:deployment}

\subsection{Predictive use case in RecSys}\label{section:recsys-deploy}
SLM variants of the RecSys use case have been deployed to a large-scale A/B test.

\noindent \textbf{Serving Infrastructure} For all use cases, we benchmark and serve traffics using nodes with 256 CPU cores, 2TB of host memory and 8 NVIDIA H100 GPUs. As discussed in Appendix~\ref{subsec:Training and serving efficiency}, we deploy SGLang (version 0.4.1) as the serving engine. We use tensor parallelism to concurrently use more than 1 GPU for inference. To maximize performance, we employ FP8 quantization for both weights and activations and use FlashInfer~\cite{ye2025flashinfer} as the primary attention backend. Moreover, SGLang incorporates RadixAttention, which enables prefix caching for prompts sharing common prefixes.

\noindent \textbf{Workloads} For the RecSys workload, we follow the prompt structure outlined in Section~\ref{section:prelim} and utilize context lengths of 16k and 32k. Given the predictive nature of the task, the output length (i.e., the number of generated tokens) is set to 1, rendering the workload heavily dependent on the prefill phase. Consequently, optimizing the prefill stage is crucial for performance. For instance, prefix caching substantially improves prefill (and decode) times when a prompt’s key (K) and value (V) tensors for its shared prefix have already been processed. In cases where there are $k$ candidate items to be ranked for a member $m$, $k$ prompts are served. These prompts share a long prefix containing the user information and historical item interactions. Once one prompt is served, its KV tensors are cached, allowing subsequent prompts for the same member to reuse the cached data - the process we refer to as a hot prefill.

\noindent \textbf{Metrics} We use two key metrics - time to first token (TTFT) and time per output token (TPOT). For prediction tasks that are prefill-intensive, TTFT is the primary metric as it reflects the duration of the prefill phase. For generative tasks, both TTFT and TPOT are important. We report the total serving throughput for various context lengths.

\begin{figure}[htbp]
  \centering
  \begin{tikzpicture}
    \begin{axis}[
      xlabel={LLMs},
      ylabel={P99 TTFT (ms)},
      xlabel style={font=\large},
      ylabel style={font=\large},
      xtick=data,             
      symbolic x coords={FM,8B,6.4B,3B,2.4B,2.1B},
      x tick label style={rotate=45, anchor=east},
      width=0.9\columnwidth,  
      height=0.7\columnwidth, 
      legend style={at={(0.48,0.82)},font=\tiny, anchor=west} 
    ]
      \addplot[
        mark=*,
        color=myblue,
        thick,
        nodes near coords={\pgfkeysvalueof{/pgfplots/y}},
        every node near coord/.append style={font=\footnotesize, anchor=south}
      ] coordinates {
        (8B,282)
        (6.4B,262)
        (3B,209)
        (2.4B,197)
        (2.1B,184)
      };
      \addlegendentry{m=1, k=1, 16k};

      \addplot[
        mark=square*,
        color=myorange,
        thick,
        nodes near coords={\pgfkeysvalueof{/pgfplots/y}},
        every node near coord/.append style={font=\footnotesize, anchor=south}
      ] coordinates {
        (8B,643)
        (6.4B,613)
        (3B,472)
        (2.4B,456)
        (2.1B,391)
      };
      \addlegendentry{m=1, k=1, 32k};

      \addplot[
        mark=triangle*,
        color=mygreen,
        thick,
        nodes near coords={\pgfkeysvalueof{/pgfplots/y}},
        every node near coord/.append style={font=\footnotesize, anchor=south}
      ] coordinates {
        (8B,671)
        (6.4B,655)
        (3B,500)
        (2.4B,488)
        (2.1B,403)
      };
      \addlegendentry{m=1, k=4, 32k};

    \end{axis}
  \end{tikzpicture}
  \caption{P99 TTFT (ms) for various LLMs}
  \label{figure:p99ttft}
\end{figure}
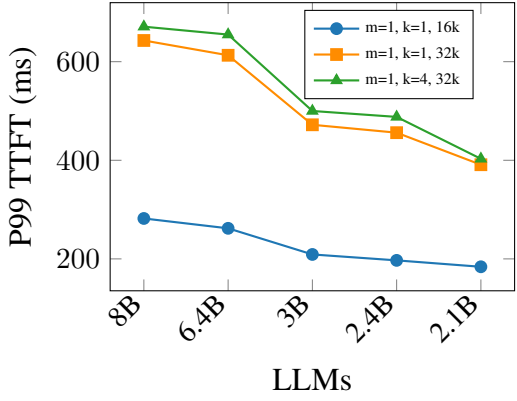

\noindent \textbf{Results} In terms of quality, the models under consideration perform similarly with comparable AUCs. For performance, we report TTFT and throughput measurements for both 16k and 32k context lengths. P99 TTFT results for a workload with 1 QPS (i.e., $m=1$) and 1 or more prompt per member (i.e., $k=1$ or more) can be found in Figure~\ref{figure:p99ttft} (detailed p50, p99 and throughput numbers can be found in Tables~\ref{table:m1k1_16k}, ~\ref{table:m1k1_32k} and ~\ref{table:m1k4_32k} in Appendix~\ref{subsec:Extra tables}). From the results, we can conclude that latency drops drastically as model size becomes smaller. Serving traffic with 32k context is significantly slower than that with 16k context. Setting $k$ more than 1 doesn't hurt latency much, because of KV caching. 

\begin{figure}[htb]
    \centering
    \begin{tikzpicture}
        \begin{axis}[
            ybar stacked,
            symbolic x coords={3B (16k), 2.1B (16k), 3B (32k), 2.1B (32k)},
            xtick=data,
            x tick label style={rotate=45, anchor=east},
            ylabel={Latency (ms)},
            ymin=0,
            legend pos=north west,
            bar width=15pt,
            x tick style={draw=none}, 
            width=\columnwidth,
            height=6cm,
            legend image code/.code={
              \draw[] (0,-0.075cm) rectangle (0.03cm,0.075cm);
            }
        ]
        
        \addplot+[ybar, 
                    fill=myblue!50!white, 
                  draw=myblue!55!white,
          ] plot coordinates {
            (3B (16k), 3)
            (2.1B (16k), 2.5)
            (3B (32k), 7)
            (2.1B (32k), 5)
        };
        
        \addplot+[ybar, 
                  fill=myorange!50!white, 
                 draw=myorange!60!white,
          ] plot coordinates {
            (3B (16k), 1)
            (2.1B (16k), 1)
            (3B (32k), 2)
            (2.1B (32k), 1.8)
        };
        
        \addplot+[ybar, 
fill=mygreen!50!white, 
                  draw=mygreen!55!white,
        ] plot coordinates {
            (3B (16k), 1)
            (2.1B (16k), 1)
            (3B (32k), 2.5)
            (2.1B (32k), 2)
        };

        \legend{Attention, AllReduce, MLP}
        \end{axis}
    \end{tikzpicture}
    \caption{Latency breakdown of a single Transformer block for pruned and unpruned models. At longer context sizes, attention is a bottleneck.}
    \label{figure:profiling}
\end{figure}
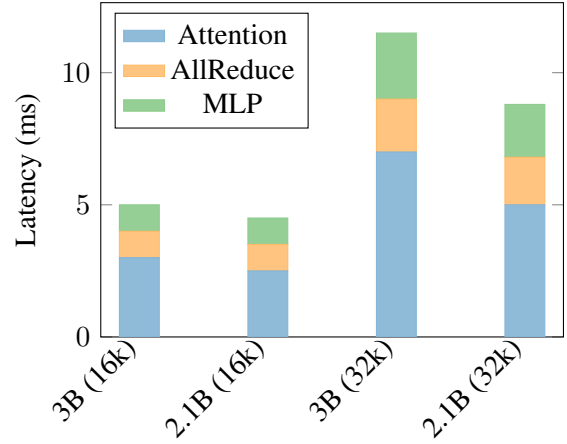

To better understand the effect of model pruning on inference latency, we present the break down of forward pass for a single layer in Figure~\ref{figure:profiling}. As it can be seen, the attention step is the main latency bottleneck. Our structured pruning of the attention heads improves the attention latency by about $40\%$ which in turn results in more than $28\%$ speed up in prefill latency.

\subsection{Generative use case in RecSys}

The reasoning task in RecSys was launched online for an 1\% A/B test. Along with data changes, KD helped the model improve by 20.29\% on an internal quality metric (IQM). We also discuss deployment lessons from a generative use case to study the effect of different quantization schemes on model inference speed and accuracy. 

\noindent \textbf{Serving Infrastructure} Our setup is mostly similar to Section~\ref{section:recsys-deploy}. However, in addition to using NVIDIA H100 GPUs, we also study the effect of using older NVIDIA A100 GPUs. To this end, 
we use the vLLM backend (version 0.6.1) for serving. We use 1 GPU for serving. 

\noindent \textbf{Workloads} The workload here consists of prompts with varying lengths, averaging to 3.8k tokens per request, with 1 request per second. The output generation is capped to 2k tokens. As we focus on a generative task here, we report both TTFT and TPOT. We study a Llama3-based model with the 8B size, and consider several serving scenarios with or without quantization using different hardware.

\noindent \textbf{Performance Results} The inference speed results are reported in Table~\ref{table:quantization}. Using the state-of-the-art H100 GPUs results in faster inference (both in terms of TTFT and TPOT) compared to A100 GPUs. In particular, we observe that FP8 serving with H100s leads to the smallest TTFT and TPOT. However, for A100 GPUs, INT4(W4A16) quantization yields the most speed-up, while INT8 (W8A8) is more appropriate for prefill-heavy tasks. For the sake of completeness, we present a brief comparison of quantization methods in terms of accuracy in Appendix~\ref{appendix:quantization}.

\begin{table}[]
\scriptsize
\centering
\begin{tabular}{cccc}
\hline \hline
Model & P50 TTFT (ms) & P50 TPOT (ms) & GPU  \\ \hline \hline
FP16    &   136 &  10.3         & H100          \\
FP8 & 122 &  9.4 & H100 \\
FP16    &       332     & 18.3 & A100          \\
W8A8 (INT) &     227       & 12.9 & A100   \\
W4A16 (INT) &       389     &  11.2 & A100   \\
\hline \hline
\end{tabular}
\caption{Comparison of different quantization methods for the Llama-3 8B model.}
\label{table:quantization}
\end{table}

\section{Limitations}

In this paper, we have studied extensively on the training and deployment of efficient SLMs for industry use cases in Recommendation Systems. We do not study any use cases beyond that; thus, our work does not cover all the techniques related to efficient LLM deployment and training. There are several limitations that we want to briefly mention here: 

1. We did not include the state-of-the-art (SoTA) sparse attention techniques due to the Inference Engine compatibility issues. Recent work such as Star Attention~\cite{starattn2024} has shown promising results in offline inference speedup. We have explored SoTA sparse attention techniques and achieved $>$3x of offline inference speedups. However, the SGLang inference engine currently does not support those techniques. We are working closely with the SGLang team to align with their roadmap, and hopefully to enable those features in the near future. 

2. For the LLM model pruning techniques discussed in this paper, we focus only on structured pruning due to its elegant theoretical guarantees and hardware efficiency. However, recent work in unstructured pruning~\cite{mlsysunstructuredsparsity,acl2024unstructuredpruning} has shown promising performance in inference speed-ups and model quality preservation. A future direction could be to combine unstructured sparsity with structured pruning to further enhance LLMs inference efficiency.  To this end, one can draw inspiration from recent work on algorithmic approaches for unstructured sparsity~\cite{meng2024alps,sun2023simple}. Other directions include exploring compression strategies adaptive to downstream parameter efficient fine-tuning~\cite{makniunified-25}.

\section{Acknowledgement} Rahul Mazumder contributed to this work while he was a consultant for LinkedIn as an Academic Scholar (in compliance
with Massachusetts Institute of Technology's outside professional activities policies).

\bibliography{reference.bib}

\appendix
\clearpage

\section{Methods}
\label{section:methods}

In this section, we detail various techniques that allow SLMs to retain strong generalization or task-specific performance, while allowing efficient serving from a latency or throughput standpoint. Specifically, we discuss training via knowledge distillation and post-training model compression. We also intersperse serving and training efficiency concerns across the entire section.

\subsection{Knowledge Distillation}

Modern LLMs work with tokens as the currency of input and output. Let $\mathbf{x}=[x_1, x_2, x_3, \dots]$ represent an input prompt consisting of a sequence of tokens. Given this prompt, a large language model (LLM) generates a response $\mathbf{y}=[y_1, y_2, y_3, \dots, y_T]$, producing tokens sequentially in an autoregressive manner. An LLM models the probability distribution $q_\theta(\mathbf{y}|\mathbf{x})$, parametrized by $\theta$.

Knowledge distillation (KD)~\cite{hinton2015distilling} transfers knowledge from a larger and expressive ``teacher'' model to a smaller ``student'' model, allowing the latter to approximate teacher performance with reduced computational resources. KD can be broadly performed in two different ways (1) by leveraging the output of a teacher model to train the student~\cite{tunstall2023zephyr,guo2025deepseek}(also known as \textbf{black-box} distillation) or (2) by leveraging intermediate outputs~\cite{muralidharan2024minitron,hinton2015distilling} (also known as \textbf{white-box} distillation). White-box distillation using the soft probabilistic outputs of the teacher is a powerful technique and helps provide richer information than hard labels used in supervised fine-tuning (SFT), helping the student generalize better, especially in tasks where smaller models struggle to discover patterns in noisy data.

We consider white-box KD using a training objective with the following general structure. Formally, given a fixed teacher model distribution $p(\mathbf{y}|\mathbf{x})$, the student model $q_\theta$ under the same vocabulary is trained by minimizing the following objective:
\begin{align}
\mathcal{L}[p_{\mathbf{y}}, \mathcal{D}(p \| q_\theta)] 
&= \\
\mathbb{E}_{\mathbf{x} \sim p_{\mathbf{x}}} \mathbb{E}_{\mathbf{y} \sim p_{\mathbf{y}}(\cdot|\mathbf{x})} \Bigg[ \sum_{t=1}^{T} \mathcal{D}\big(p(\cdot|\mathbf{y}_{<t}, \mathbf{x}) \notag & \| q_\theta(\cdot|\mathbf{y}_{<t}, \mathbf{x})\big) \Bigg]
\end{align}
where $p_{\mathbf{y}}$ denotes the distribution from which the response $\mathbf{y}$ is sampled, $\mathcal{D}$ is a divergence measure between two next-token distributions, and $T$ is the maximum response length. This objective emphasizes two aspects: 
\begin{enumerate}
  \item Responses are drawn from $p_{\mathbf{y}}$, which may correspond to ground truth data, the teacher model (sequence-level KD) ~\cite{kim2016sequence}, or the student model itself (on-policy KD) ~\cite{agarwal2024policy,gu2024minillm,zhou2023distillspec}. Recent advancements~\cite{xu2024speculative,ko2024distillm} explore a balance between on-policy and off-policy sampling to mitigate the mismatch between student-generated responses and the teacher’s distribution while addressing inefficiencies in online student autoregressive training.
  \item The student model is optimized to minimize the discrepancy between its next-token distribution $q_\theta$ and the teacher’s predictions $p$, ensuring knowledge transfer across the response sequence.
\end{enumerate}

In this work, we explore different KD strategies based on the task requirements. We also experiment with various student initialization techniques and divergence measures.

Let $\mathcal{V}$ denote the vocabulary. The commonly used divergences are:
\begin{itemize}
    \item Forward Kullback-Leibler (KL) Divergence (FKL):
\begin{align*}
&\mathcal{D}_{\text{FKL}} \left[ p(y_t | \mathbf{y}_{<t}, \mathbf{x}) \| q_\theta (y_t | \mathbf{y}_{<t}, \mathbf{x}) \right] \\ & = \sum_{i \in \mathcal{V}} p(i | \cdot) \log \left( \frac{p(i | \cdot)}{q_\theta (i | \cdot)} \right),
\end{align*}
    \item Reverse Kullback-Leibler Divergence (RKL):
\begin{align*}
& \mathcal{D}_{\text{RKL}} \left[ p(y_t | \mathbf{y}_{<t}, \mathbf{x}) \| q_\theta (y_t | \mathbf{y}_{<t}, \mathbf{x}) \right] \\ & = \sum_{i \in \mathcal{V}} q_\theta (i | \cdot) \log \left( \frac{q_\theta (i | \cdot)}{p(i | \cdot)} \right),
\end{align*}
    \item Jensen-Shannon Divergence (JSD):
\begin{align*}
& \mathcal{D}_{JS(\beta)} \left[ p(y_t | \mathbf{y}_{<t}, \mathbf{x}) \| q_\theta(y_t | \mathbf{y}_{<t}, \mathbf{x}) \right] \\ & = \;    \beta \mathcal{D}_{\text{FKL}} \left[ p \| m \right] + (1 - \beta) \mathcal{D}_{\text{FKL}} \left[ q_\theta \| m \right],
\end{align*}
where \( m = \beta p + (1 - \beta) q_\theta \) is the mixture distribution.
\end{itemize}

\subsection{Post-training model compression}\label{sec:appendix-compression}

Model compression is a widely studied area of machine learning. We specifically focus on post-training compression (PTC) techniques for improving the inference efficiency of LLMs.

In post-training compression, we apply compression to the model after training. A common compression procedure is based on pruning or quantizing the model weights of a pre-trained model but it can result in large loss in model utility due to which alternative approaches are preferred. 
Specifically, recent compression procedures employ a layerwise approach (Equation~\ref{eq:layerwise}) where the utility for every layer in a model is retained (to the extent possible) by minimizing a suitable layerwise objective function based on a calibration dataset --- this approach can be scaled to large models while retaining model utility, and we use this method in our work.

We describe a mathematical framework for layerwise PTC using calibration data. Let
$\mathbf{X} \in \mathbb{R}^{n \times d}$
denote the calibration data that serve as inputs to a linear layer of the model (e.g., an MLP or an attention projection). Here, \( n \) is the number of tokens in the calibration dataset, and \( d \) is the input dimension of the layer. For instance, in the case of an MLP down projection layer of a Transformer block, \( d \) corresponds to the intermediate size of the model.

Furthermore, let $\mathbf{W} \in \mathbb{R}^{d \times p}$
denote the weight matrix of the layer, where \( p \) is the output dimension. In the MLP down projection example, \( p \) represents the hidden size of the model. We denote by
$\hat{\mathbf{W}} \in \mathbb{R}^{d \times p}$
the weight matrix after compression. The layerwise reconstruction error is defined as
$\|\mathbf{X}\mathbf{W} - \mathbf{X}\hat{\mathbf{W}}\|_F^2.$
Thus, for each layer that undergoes compression, we consider an optimization problem of the form
\begin{equation}\label{eq:layerwise}
\min_{\hat{\mathbf{W}}} \|\mathbf{X}\mathbf{W} - \mathbf{X}\hat{\mathbf{W}}\|_F^2 \quad \text{subject to} \quad \hat{\mathbf{W}} \in \mathcal{Q},
\end{equation}
where \( \mathcal{Q} \subseteq \mathbb{R}^{d \times p} \) denotes the set of feasible solutions that conform to a particular compression scheme (eg, unstructured or structured pruning, quantization, etc), In practice, the set \( \mathcal{Q} \) often exhibits a discrete structure, which renders the optimization problem in \eqref{eq:layerwise} challenging to solve. Past work has shown that a better optimization procedure for Problem~\eqref{eq:layerwise} generally results in better utility-compression tradeoffs~\cite{meng2024osscar,meng2024alps,
meng2024falcon,behdin2023quantease}, which motivates the approaches we used in our work. 

Here, we consider two compression techniques: \\
\noindent\textbf{Quantization} In quantization, the model weights are represented in lower precision, using a fewer number of bits. Quantization has proved to be successful in the LLM domain, to obtain compressed models with small accuracy loss~\cite{frantar2022gptq,smoothquant}. In this work, we consider weight-only quantization, where only model weights are quantized, as well as weight and activation quantization. We study methods such as GPTQ~\cite{frantar2022gptq} and QuantEase~\cite{behdin2023quantease} for 4-bit weight-only quantization (aka W4A16), SmoothQuant~\cite{smoothquant} for 8-bit weight-and-activation quantization (aka W8A8), and 8-bit floating-point (FP8) quantization. Since quantization is dependent on hardware, we discuss the details of quantization-related experiments in Section~\ref{section:deployment} (deployment).

\noindent\textbf{Structured Pruning} Post-training neural network pruning has 
old roots~\cite{obd,obs} -- the basic idea is to identify ``redundant'' model weights and set them to zero to reduce model footprint.  
Recently, due to increasing challenges associated with large models, advanced algorithms have been explored for pruning neural networks at the post-training stage.  
Some unstructured pruning methods for LLMs using layerwise reconstruction error include~\citet{meng2024alps,frantar2023sparsegpt,sun2023simple}. Other pruning approaches using loss functions different from layerwise reconstruction error include approaches based on global Fisher loss~\cite[and references therein]{benbaki2023fast,meng2024falcon}, layerwise loss functions that depend upon the future layers~\cite{lucaspreserving-24}, and low-rank fine-tuning adapted loss functions~\cite{makniunified-25}[see references therein] --- we leave the exploration of these approaches as interesting directions for future research. 

Without any structure on the model sparsity, however, it can be difficult to realize any inference acceleration from pruning\footnote{Some of the approaches discussed 
above apply~\cite{meng2024alps,sun2023simple} to semi-structured pruning such as 2:4 sparsity which is different from structured pruning.}. Therefore, in this work, we pursue a structured pruning approach.
In structured pruning, the goal is to obtain smaller models via removing some neurons from the model weights. In particular, we study MLP pruning, where the goal is to reduce the intermediate size of the model via removing some hidden neurons in feed-forward layers. We also study attention pruning, where we remove a certain number of attention heads from the model~\cite{meng2024osscar,structured1,structured2}.
In this paper, we use OSSCAR~\cite{meng2024osscar} which uses a discrete optimization approach for structured pruning. OSSCAR can be scaled to the large-scale problems we consider here. We use OSSCAR as it can result in state-of-the-art performance when it comes to post-training structured pruning of LLMs and leads to the least drop in accuracy when compared to other methods.

\subsection{Training and serving efficiency} 
\label{subsec:Training and serving efficiency}

Despite significant algorithmic advances, the challenges of training and serving LLMs persist. Efficient training and serving remain critical for practical deployment, requiring ongoing improvements in kernel optimization, distributed training, and inference acceleration.

\noindent
\textbf{Training Efficiency} LLM training presents a formidable challenge due to the sheer scale of these models and the quadratic complexity of transformer architectures. Model FLOPs utilization (MFU)~\cite{chowdhery2023palm} is commonly used to measure GPU efficiency, making it necessary to optimize kernel operations and distributed training strategies. We have implemented Liger Kernel~\cite{hsu2024liger} in Triton~\cite{tillet2019triton}, incorporating several key optimizations. First, we employ kernel fusion to reduce repetitive memory transfers between SRAM and DRAM. Next, we adopt in-place tensor modifications to avoid creating additional tensors whenever possible, thus lowering the memory footprint. We also apply chunking, which prevents the full materialization of large tensors and provides tuning flexibility while maintaining comparable performance. In combination, these optimizations reduce training time by 20\% and memory usage by 60\%. Additional performance-memory tradeoffs such as gradient checkpointing and CPU offloading can lead to as much as a threefold speedup.

For distributed training, we use ZeRO~\cite{rajbhandari2020zero} to shard model parameters and data across multiple GPUs, overlapping computation with communication to sustain high MFU. Together with the DeepSpeed team, we have optimized the ZeRO algorithm, and for network-constrained clusters, we have developed ZeRO++~\cite{wang2023zero++,dai2024enhancing} to mitigate non-deterministic synchronization issues that can hinder convergence. ZeRO++ provides a 2.4$\times$ speedup over vanilla ZeRO.

\noindent
\textbf{Serving Efficiency} Serving LLMs efficiently poses unique challenges due to both high computational demands and strict latency requirements. In production environments, the choice of serving frameworks is pivotal for maximizing throughput and minimizing response times. Several solutions, including vLLM, SGLag, TRT-LLM, and MLC-LLM~\cite{kwon2023vllm,zheng2024sglang,TRT-LLM,mlc-llm}, have been proposed to address these needs. In our use cases, we extensively evaluated vLLM and SGLang. Our benchmarks revealed that SGLang is better suited to our workloads because its radix tree-based caching mechanism aligns well with our usage patterns and it integrates tightly with FlashInfer~\cite{ye2025flashinfer}, whose efficient attention kernels accelerate the sequence lengths and batch sizes we typically handle.

To further improve serving performance, we deploy our models on NVIDIA H100 GPUs at FP8 precision, striking a practical balance between computational efficiency and model quality. Additional details regarding our LLM serving engine configurations can be found in Section~\ref{section:deployment}.

\section{Additional Numerical Experiments}
\subsection{Comparison of Quantization Methods}\label{appendix:quantization}

We present a comparison of different quantization methods. We use the Meta Llama 3.1 8B Instruct model, and quantize the model using 1024 samples from the open source C4~\cite{raffel2020exploring} dataset as the calibration set. We report the zero-shot accuracy of the model on three open-source tasks PIQA~\cite{piqa} and ARC easy/challenge~\cite{arc}. We compare W8A8 quantization with SmoothQuant~\cite{smoothquant}, FP8 quantization on H100 GPUs, and W4A16 quantization with GPTQ~\cite{frantar2022gptq} and QuantEase~\cite{behdin2023quantease}.
The results are shown in Table~\ref{table:quantization-accuracy}. We see that 8-bit quantization generally has a small loss of accuracy. On the other hand, GPTQ with W4A16 shows some model quality degradation. However, using QuantEase for better optimization helps to reduce the model quality gap. In our internal experiments, we have observed similar trends when comparing different methods.

\begin{table}[]\scriptsize\centering
\begin{tabular}{cccc}
\hline \hline
Model & ARC-c & ARC-e & PIQA \\ \hline \hline
FP16    & 0.5299   &     0.8136      &    0.7982   \\
FP8    &  0.5179  &     0.8056      &    0.7922      \\
W8A8-INT    &  0.5171  &     0.8123      &     0.7954      \\
W4A16-INT-GPTQ    &   0.436 &      0.7306     &      0.7437    \\
W4A16-INT-QuantEase    & 0.5077   &         0.8068  &      0.7954     \\
\hline \hline
\end{tabular}
\caption{Comparison of different quantization schemes with the Llama 3.1 8B Instruct model.}
\label{table:quantization-accuracy}
\end{table}

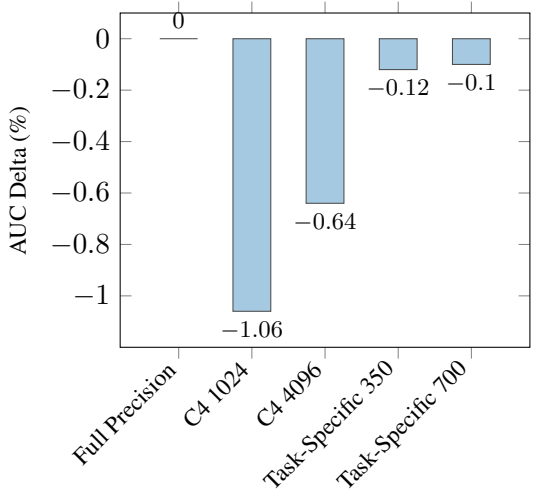
\begin{figure}[t]
    \centering
    \begin{tikzpicture}
        \begin{axis}[
            ybar,
            bar width=0.5cm,
            width=7cm,
            height=6cm,
            ylabel={AUC Delta (\%)},
            ylabel style={font=\small},
            symbolic x coords={Full Precision,C4 1024,C4 4096,Task-Specific 350,Task-Specific 700},
            xtick=data,
            enlarge x limits=0.2,
            ymin=-1.2,  
            ymax=0.1,
            ytick={-1.0, -0.8, -0.6, -0.4, -0.2, 0.0},
            nodes near coords,
            every node near coord/.append style={font=\footnotesize},
            x tick label style={font=\small, rotate=45, anchor=east},
            title style={font=\small},
        ]
        \addplot[fill=myblue!40!white, draw=black!70] coordinates {
            (Full Precision, 0.00)
            (C4 1024,      -1.06)
            (C4 4096,      -0.64)
            (Task-Specific 350, -0.12)
            (Task-Specific 700, -0.10)
        };
        \end{axis}
    \end{tikzpicture}
    \caption{Comparison of one-shot pruning methods. The bars indicate the drop (in percentage points) relative to the full precision baseline. The pruned model is a 6.4B model (20\% MLP pruning).}
    \label{fig:calibration}
\end{figure}

\subsection{Effect of calibration data for Pruning}
\label{appendix:calibration_data_pruning}
Figure~\ref{fig:calibration} captures the effect of the calibration dataset $\boldsymbol{X}$ (discussed in Section~\ref{section:methods} and Equation~\eqref{eq:layerwise}) on the accuracy of the pruned versions of the 8B student model (results for the 3B model are similar and are hence omitted for clarity). The \emph{Full Precision} bar illustrates the baseline accuracy of the non-pruned model. We consider two different datasets for calibration - C4~\cite{raffel2020exploring}, an open source dataset, and an in-domain dataset. When we prune using a randomly sampled portion of the C4 dataset (1,024 or 4,096 examples), accuracy drops, although more samples mitigate this drop to an extent. These results indicate that increasing the number of calibration examples from 1,024 to 4,096 can partially recover lost accuracy due to pruning. However, leveraging fewer but more domain-specific samples (350 or 700 examples from the target task) yields better accuracy values, which closely match the full precision baseline. This highlights the importance of using task-relevant data for calibration, even if it involves fewer examples, as it can produce more accurately pruned models than generic calibration sets.

\subsection{Extra tables for Section~\ref{section:deployment}}
\FloatBarrier
\label{subsec:Extra tables}
\begin{table}[htbp]\scriptsize\centering
\begin{tabular}{llll}
\hline\hline
Model & P50 TTFT (ms) & P99 TTFT (ms) & Throughput \\ \hline\hline
FM    & 1032           & 1039           & 14127       \\
8b    & 271           & 282           & 14121      \\
6.4B  & 256           & 269           & 14121      \\
3B    & 195           & 209           & 14121      \\
2.4B  & 189           & 197           & 14122      \\
2.1B  & 171           & 184           & 14110      \\  \hline\hline
\end{tabular}
\caption{Results for $m=1$, $k=1$ for $16k$ context length using 4 GPUs (tp=4).}
\label{table:m1k1_16k}
\end{table}

\begin{table}[htbp]\scriptsize\centering
\begin{tabular}{llll}
\hline\hline
Model & P50 TTFT (ms) & P99 TTFT (ms) & Throughput  \\ \hline\hline
FM    & 407661           & 45791           & 15740   \\
8b    & 626           & 643           & 28427       \\
6.4B  & 600           & 613           & 28427       \\
3B    & 452           & 472           & 28423       \\
2.4B  & 437           & 456           & 28422      \\ 
2.1B  & 367           & 391           & 28420       \\ \hline\hline

\end{tabular}
\caption{Results for $m=1$, $k=1$ for $32k$ context length using 4 GPUs (tp = 4).}
\label{table:m1k1_32k}
\end{table}

\begin{table}[htbp]\scriptsize\centering
\begin{tabular}{llll}
\hline \hline
Model & P50 TTFT (ms) & P99 TTFT (ms) & Throughput  \\ \hline \hline
FM    & 179483           & 370376           & 45081       \\ 
8b    & 646           & 671           & 115568      \\ 
6.4B  & 626           & 655           & 115560      \\ 
3B    & 477           & 500           & 115546      \\ 
2.4B  & 465           & 488           & 115544      \\ 
2.1B  & 378           & 403           & 115520      \\ \hline \hline
\end{tabular}
\caption{Results for $m=1$, and $k=4$ for $32k$ context length using 4 GPUs (tp=4).}
\label{table:m1k4_32k}
\end{table}

\section{Implementation and Efficiency Checklist}
We provide an open source implementation of all methods discussed in this paper, in addition to examples to recreate our distillation and pruning pipelines. Our python package can be found at \href{https://github.com/linkedin/FMCHISEL}{https://github.com/linkedin/FMCHISEL}.

Additionally, we summarize the lessons from our experiments by providing a practical checklist to create efficient SLMs through distillation and pruning for a specific task. We assume one has access to a family of pre-trained LLMs such as Llama.

\begin{enumerate}
    \item Create a teacher model, for example, by fine-tuning a large pre-trained on the desired task(s).
    \item Perform distillation using the teacher model developed above, and a variety of student models.
    \item Find the smallest student model that meets the desired quality bar.
    \item Perform profiling on the selected student model to identify inference bottlenecks.
    \item Prune and distill the selected student model, targeting bottlenecks identified above, until the pruned model's quality approaches the quality bar. This is the final model that will be deployed.
    \item Perform extensive benchmarking under various workloads and scenarios (such as hardware failure, bursty traffic, etc.) to obtain a conservative estimation of the number of GPUs required.
\end{enumerate}

\section{Ethical Concerns}
This paper presents a methodological study of model compression techniques such as model pruning and knowledge distillation. These techniques aim to ensure the compressed models produce similar predictions compared to the original ones and hence, these methods do not pose any inherent ethical concerns. The analysis of larger (teacher) models is out of the scope of this paper.

\end{document}